\documentstyle[12pt]{article}

\begin{document}
\setcounter{page}{1} \pagestyle{plain} \vspace{1cm}
\begin{center}
\Large{\bf On the cosmological viability of the Hu-Sawicki type modified induced gravity }\\
\small \vspace{1cm}
{\bf Kourosh Nozari\footnote{knozari@umz.ac.ir}}\quad\ and \quad {\bf Faeze Kiani\footnote{fkiani@umz.ac.ir}}\\
\vspace{0.5cm} {\it Department of Physics,
Faculty of Basic Sciences,\\
University of Mazandaran,\\
P. O. Box 47416-95447, Babolsar, IRAN}\\
\end{center}
\vspace{1.5cm}
\begin{abstract}

It has been shown recently that the normal branch of a DGP
braneworld scenario self-accelerates if the induced gravity on the
brane is modified in the spirit of $f(R)$ modified gravity. Within
this viewpoint, we investigate cosmological viability of the
Hu-Sawicki type modified induced gravity. Firstly, we present a
dynamical system analysis of a general $f(R)$-DGP model. We show
that in the phase space of the model, there exist three standard
critical points; one of which is a de Sitter point corresponding to
accelerating phase of the universe expansion. The stability of this
point depends on the effective equation of state parameter of the
curvature fluid. If we consider the curvature fluid to be a
canonical scalar field in the equivalent scalar-tensor theory, the
mentioned de Sitter phase is unstable, otherwise it is an attractor,
stable phase. We show that the effective equation of state parameter
of the model realizes an effective phantom-like behavior. A
cosmographic analysis shows that this model, which admits a stable
de Sitter phase in its expansion history, is a cosmologically viable
scenario.
\end{abstract}
\vspace{1.5cm}
\newpage

\section{Introduction}

The late-time accelerating phase of the universe expansion which is
supported by data related to the luminosity measurements of high red
shift supernovae [1], measurements of degree-scale anisotropies in
the cosmic microwave background (CMB) [2] and large scale structure
(LSS) [3], is one of the challenging problems in the modern
cosmology. The rigorous treatment of this phenomenon can be provided
essentially in the framework of general relativity. In the
expression of general relativity, late time acceleration can be
explained either by an exotic fluid with large negative pressure
that is dubbed as \textit{dark energy} in literature, or by
modifying the gravity itself which is dubbed as \textit{dark
geometry} or \textit{dark gravity} proposal. The first and simplest
candidate of dark energy is the cosmological constant, $\Lambda$
[4]. But, there are theoretical problems associated with it, such as
its unusual small numerical value (the fine tuning problem), no
dynamical behavior and even its unknown origin [5]. These problems
have forced cosmologists to introduce alternatives in which dark
energy evolves during the universe evolution. Scalar field models
with their specific features provide an interesting alternative for
cosmological constant and can reduce the fine tuning and coincidence
problems. In this respect, several candidate models have been
proposed: quintessence scalar fields [6], phantom fields [7] and
Chaplygin gas [8] are among these candidates. Nevertheless, we
emphasize that the scalar field models of dark energy are not free
of shortcomings.

As an alternative for dark energy, modification of gravity can be
accounted for the late time acceleration. Among the most popular
modified gravity scenarios which may successfully describe the
cosmic speed-up, is $f(R)$ gravity [9,10]. Modified gravity also can
be achieved by extra-dimensional theories in which the observable
universe is a 4-dimensional brane embedded in a five-dimensional
bulk. The Dvali-Gabadadze-Porrati (DGP) model is one of the
extra-dimensional models that can describe late-time acceleration of
the universe in its self-accelerating branch due to leakage of
gravity to the extra dimension [11,12].

Recent observations constrain the equation of state parameter of the
dark energy to be $w_{X}\approx-1$ and even $w_{X}<-1$ [13]. One of
the candidates for dark energy of this kind is the phantom scalar
field. This component has the capability to create the mentioned
acceleration and its behavior is extremely fitted to observations.
But it suffers from theoretical problems; it violates the null
energy condition and its energy density increases with expansion of
the universe leading to a future big rip singularity. Also it causes
the quantum vacuum instabilities. So, some authors have attempted to
realize a kind of phantom-like behavior ($w_{eff}<-1$) in the
cosmological models without introduction of phantom fields. In fact,
the possibility of realization of an effective phantom nature
without introduction of phantom fields is an important task and has
been appreciated sufficiently in recent years [14].

In the streamline of the mentioned issues, we are going to study
cosmological viability of a class of DGP-inspired braneworld models
in which the induced gravity on the normal branch is modified in the
spirit of $f(R)$ gravity [10,15,16,17]. Firstly, we study the
cosmological dynamics of this model within a dynamical system
approach. We show that there exists a standard de Sitter point in
the phase plane of the model. In this respect, this model has the
potential to explain accelerated expansion of the universe. The
stability of this point depends completely on the effective equation
of state parameter of the curvature fluid. If we consider the
curvature fluid to be a canonical scalar field in the equivalent
scalar-tensor theory, the mentioned de Sitter phase is unstable,
otherwise it is an attractor, stable phase. Since the late-time
accelerating phase of the universe expansion is explained by a
stable de Sitter phase, we can investigate the cosmological
viability of such theoretical models based on the phantom-like
behavior of this $f(R)$-DGP gravity. To be more specific, in which
follows we focus on the cosmological viability of the Hu-Sawicki
type modified induced gravity and show that this model has
capability to realize a stable, attractor de Sitter phase. We point
out that the phantom mimicry discussed in this study has a geometric
origin. To be more realistic, we compare our results with
observation via a cosmographic approach.

\section{$f(R)$-DGP scenario}

In this section, possible modification of the induced gravity on the
brane is investigated in the spirit of $f(R)$ theories
[10,15,16,17]. It has been shown that $4D$ $f(R)$ theories in the
present time can follow closely the expansion history of the
$\Lambda$CDM universe [18]. Here we study an extension of $f(R)$
theories to a DGP braneworld setup. The motivation behind this study
is that modified induced gravity on the normal branch of a DGP
scenario provides some new interesting features, one of which is
self-acceleration of the normal DGP branch in this situation (see
Refs. [10,16,17] for details). Similar to the normal branch of the
standard DGP cosmology, the resulting generalized normal branch is
also ghost-free and therefore the issue of ghost-instabilities is
irrelevant in this case [17]. The action of this model can be
written as follows
\begin{equation}
{\cal{S}}=\frac{M_{5}^{3}}{2}\int d^{5}x\sqrt{-g} {\cal{R}} + \int
d^{4}x\sqrt{-q}\Big(M_{5}^{3}\overline{K}+{\cal{L}}\Big)\,,
\end{equation}
where by definition
\begin{equation}
{\cal{L}}=\frac{m_{p}^{2}}{2}f(R)+{\cal{L}}_{m}\,.
\end{equation}
By calculating the bulk-brane Einstein's equations and using a
spatially flat FRW line element, the following modified Friedmann
equation is obtained [15,16,17]
\begin{equation}
H^{2}=\frac{8\pi
G}{3}\bigg(\rho^{(m)}+\rho^{(rad)}+\rho^{(curv)}\bigg)\pm\frac{H}{\bar{r}_{c}}
\end{equation}
where
\begin{equation}
\rho^{(curv)}=m_{p}^{2}\bigg(\frac{1}{2} \Big[f(R)-R f'(R)\Big]
-3\dot{R}H f''(R) \bigg)\,,
\end{equation}
is energy density corresponding to the curvature part of the theory.
This energy density can be dubbed as \emph{dark curvature} energy
density. $\bar{r}_{c}$ is the re-scaled crossover distance that is
defined as $\bar{r}_{c}=r_{c}f'(R)$ and a prime marks
differentiation with respect to the Ricci scalar, $R$. We note that
in this scenario there is an effective gravitational constant, which
is re-scaled by $f'(R)$ so that $G=G_{eff}\equiv\frac{1}{8\pi
m_{p}^{2}f'(R)}$ [15]. In order to study the phase space of this
scenario, it is more suitable to rewrite the normal branch of the
Friedmann equation (3) in the following more phenomenological form
\begin{equation}
E^{2}=\Omega_{m}(1+z)^3+\Omega_{rad}(1+z)^4+
\Omega_{curv}(1+z)^{3(1+w_{curv})}-2\sqrt{\Omega_{r_{c}}}E\,,
\end{equation}
where by definition $$\Omega_{curv}=\frac{8\pi
G}{3H_{0}^{2}}\rho_{0}^{(curv)}\,\,,\quad\quad
\Omega_{r_{c}}=\frac{1}{4[r_{c}f_{0}'(R)]^{2}H_{0}^{2}}\,,$$ and
also

\begin{equation}
w_{curv}=-1+\frac{\ddot{R}f''(R)+\dot{R}\Big[\dot{R}f'''(R)-
Hf''(R)\Big]}{\frac{1}{2}[f(R)-Rf'(R)]-3H\dot{R}f''(R)}\,.
\end{equation}
We note that $w_{curv}$ is not a constant and varies with redshift.

\section{The phase space of a general $f(R)$-DGP model}

To investigate cosmological dynamics of this model within a
dynamical system approach, we express the cosmological equations in
the form of an autonomous, dynamical system. For this purpose, we
define the following normalized expansion variables
\begin{equation}
p=\frac{\sqrt{\Omega_{m}}}{a^{3/2}E}\,,\quad\quad
r=\frac{\sqrt{\Omega_{rad}}}{a^{2}E}\,,\quad\quad
s=\frac{\sqrt{\Omega_{curv}}}{a^{3(1+w_{curv})/2}E}\,,\quad\quad
u=\frac{\sqrt{\Omega_{r_c}}}{E}\,.
\end{equation}
In this way, equation (5) with minus sign (corresponding to the
generalized normal DGP branch) and in a dimensionless form, is
written as follows
\begin{equation}
1+2u=p^{2}+r^{2}+s^{2}\,.
\end{equation}
This constraint means that the allowable phase space of this
scenario in the $p$-$r$-$s$ space is outside of a sphere with radius
$1$, which is defined as $p^{2}+r^{2}+s^{2}\geq 1$\,. The autonomous
system is obtained as follows
\begin{equation}
\left\{\begin{array}{ll}
p'=\frac{3p\Big[p^{2}+(1+2w_{curv})r^{2}+\frac{5}{3}s^{2}-1\Big]}{2(p^{2}+r^{2}+s^{2}+1)}\,,\vspace{0.5cm}\\
r'=\frac{3r\Big[2p^{2}+\frac{8}{3}s^{2}+(1+w_{curv})(r^{2}-p^{2}-s^{2}-1)\Big]}{2(p^{2}+r^{2}+s^{2}+1)}\,,\vspace{0.5cm}\\
s'=\frac{s\Big[2s^{2}+p^{2}+(1+3w_{curv})r^{2}-2\Big]}{p^{2}+r^{2}+s^{2}+1}\,.
\end{array}\right.
\end{equation}
Here a prime marks differentiation with respect to the new time
variable $\tau=\ln{a}$ that $a$ is scale factor of the universe. The
critical points in the phase plane are obtained by solving the
equations $p'=0$, $r'=0$ and $s'=0$, that is, setting the autonomous
system (9) to be vanishing. The results are shown in table 1. To
investigate the stability of these points, we apply the linear
approximation analysis to achieve the \textit{Jacobian} matrix. Note
that the critical points and their stability depend on the value of
$w_{curv}$\,. Here we investigate the stability of the standard
critical points in two different subspaces of the model parameter
space where EoS of the curvature fluid has either a phantom or a
quintessence character. As we see in table 1, the radiation
dominated phase (point $A$) and matter dominated phase (point $B$)
in this scenario, are unstable phases of the universe expansion
independent on the value of $w_{curv}$\,. Whereas, the accelerating
phase of the universe expansion (point $C$) is a stable phase if the
curvature fluid is considered to be a non-canonical (phantom) scalar
field ($w_{curv}<-1$) in equivalent scalar-tensor theory; otherwise
it is an unstable phase. It is necessary to mention that whenever
$w_{curv}=-1$, the variables $s$ and $u$ are not independent and the
phase space is $2$D (here the curvature fluid plays the role of a
cosmological constant, the same as $\Lambda DGP$ model. For more
details see Ref. [19]). Figure 1 shows the $3$D phase space
trajectories of the model. In this figure, the point $C$ as a de
Sitter point is an attractor for $w_{curv}<-1$. Therefore, a model
universe which is described by modified induced gravity on the
normal DGP branch, has a stable, positively accelerated expansion
phase if the modified gravity indicates a phantom-like behavior. We
note that points $O$ and $C'$ do not belong to physical phase space
of our model universe.

After exploration of the cosmological dynamics in a general
$f(R)$-DGP setup within a phase space analysis, in the next section
we study cosmological viability of an specific $f(R)$-DGP model.

\begin{table}
\begin{center} \caption{Eigenvalues and the stability properties of the
critical points.} \vspace{0.5 cm}
\begin{tabular}{c c c c c c  }
  \hline
  \hline    points & $(p,r,s)$ &character& eigenvalues &  $w_{curv}<-1$ &  $w_{curv}> -1$\\

  \hline
  \\
         $A$&$(0,0,1)$& radiation &$\Big[2\,,\frac{1}{2}(1-3w_{curv})\,,\frac{1}{2}\Big] $& unstable & unstable \\
         \\
         $B$ &$(1,0,0)$& matter & $\Big[-\frac{1}{2}\,,-\frac{3}{2}w_{curv}\,,\frac{3}{2}\Big]$& unstable & unstable \\
         \\
         $C$ &$(0,1,0)$& de Sitter &$\Big[\frac{3w_{curv}-1}{2}\,,\frac{3(1+w_{curv})}{2}\,,\frac{3(1+4w_{curv})}{8}$\Big]& stable & unstable \\
         \\
     \hline
\end{tabular}
\end{center}
\end{table}

\begin{figure}[htp]
\begin{center}\includegraphics{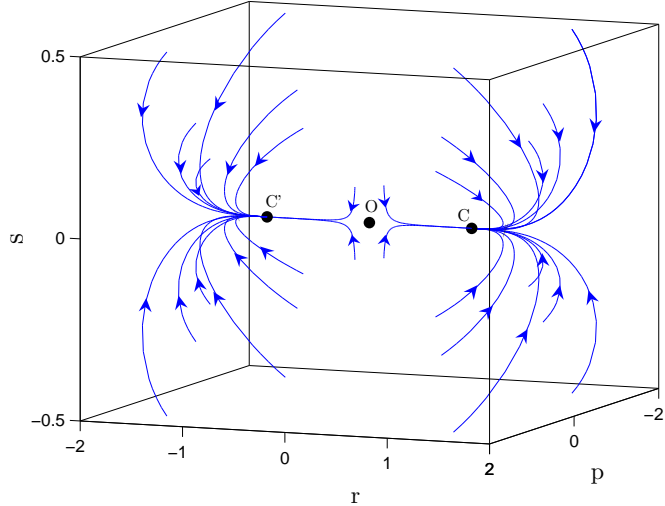} \vspace{5cm}
\end{center}
 \caption{\small {The $3$D phase space of the autonomous system (9) for $w_{curv}<-1$. There are three critical
 points: $C$ is an attractor de Sitter point, $O$ is a saddle point and $C'$ is an attractor
 point.}}
\end{figure}

\section{Cosmological viability of the Hu-Sawicki type modified
induced gravity}

Now we focuss on the cosmological viability of the model by
considering a Hu-Sawicki type modified induced gravity on the DGP
brane. It is shown in the Ref. [18] that the expansion history of
the mentioned model in $4$ dimensions is widely close to the
$\Lambda$CDM model in the high-redshift regime. Now in a braneworld
extension, we expect the Hu-Sawicki induced gravity mimics the
$\Lambda$DGP model in the mentioned regime. In other words, in this
regime curvature term in the Friedmann equation is close to the
cosmological constant which is screened by the term
$\frac{H}{\bar{r}_{c}}$\,. In fact, the dynamical screening effect
is the main origin of the phantom-like behavior of the curvature
term in the normal branch of this DGP-inspired braneworld scenario
[15]. The Hu-Sawicki model [18] is given by
\begin{equation}
f(R)=R-m^{2}\,\frac{c_{1}(\frac{R}{m^{2}})^{n}}{c_{2}(\frac{R}{m^{2}})^{n}+1},
\end{equation}
where $m^{2}$, $c_{1}$, $c_{2}$ and $n$ are free positive parameters
that can be expressed as functions of density parameters. Now we
explore the dependence of these parameters on density parameters
defined in our setup. Variation of the action (1) with respect to
the metric yields the induced modified Einstein equations on the
brane
\begin{equation}
G_{\alpha\beta}=\frac{1}{M_{5}^{6}}{\cal{S}}_{\alpha\beta}-{\cal{E}}_{\alpha\beta}\,,
\end{equation}
where ${\cal{E}}_{\alpha\beta}$ (which we neglect it in our
forthcoming arguments), is the projection of the bulk Weyl tensor on
the brane
\begin{equation}
{\cal{E}}_{\alpha\beta}=^{(5)}C_{RNS}^{M}\,n_{M}n^{R}g_{\alpha}^{N}g_{\beta}^{S}
\end{equation}
and ${\cal{S}}_{\alpha\beta}$ as the quadratic energy-momentum
correction into Einstein field equations is defined as follows
\begin{equation}
{\cal{S}}_{\alpha\beta}=-\frac{1}{4}
\tau_{\alpha\mu}\tau^{\mu}_{\beta}+\frac{1}{12}\tau\tau_{\alpha\beta}+
\frac{1}{8}g_{\alpha\beta}\tau_{\mu\nu}\tau^{\mu\nu}-\frac{1}{24}g_{\alpha\beta}\tau^{2}\,.
\end{equation}
$\tau_{\alpha\beta}$ as the effective energy-momentum tensor
localized on the brane is defined as [10]
\begin{equation}
\tau_{\alpha\beta}=-m_{p}^{2}f'(R)G_{\alpha\beta}+\frac{m_{p}^{2}}{2}\Big[f(R)-R
f'(R)\Big]g_{\alpha\beta}+T_{\alpha\beta}+m_{p}^{2}\Big[\nabla_{\alpha}\nabla_{\beta}f'(R)-g_{\alpha\beta}\Box
f'(R)\Big]\,.
\end{equation}
The trace of Eq. (11), which can be interpreted as the equation of
motion for $f'(R)$\,, is obtained as
\begin{equation}
R=\frac{5}{24M_{5}^{6}}\tau^{2}\,.
\end{equation}
$\tau$, the trace of the effective energy-momentum tensor localized
on the brane is expressed as
\begin{equation}
\tau=m_{p}^{2}\Big[2f(R)-Rf'(R)-3\Box
f'(R)\Big]-(\rho_{m}+\rho_{rad})\,,
\end{equation}
To highlight the DGP character of this generalized setup, we express
the results in terms of the DGP crossover scale defined as
$r_{c}=\frac{m_{p}^{2}}{2M_{5}^{3}}$. So, the equation of motion for
$f'(R)$ is rewritten as follows
$$\frac{5}{6}r_{c}^{2}\Bigg (\Big[2f(R)-Rf'(R)\Big]^{2}+9\Big(\Box
f'(R)\Big)^{2}+6Rf'(R)\Box f'(R)-12f(R)\Box f'(R)\Bigg)$$
\begin{equation}
+\frac{5}{6}\frac{r_{c}}{M_{5}^{3}}\Big[Rf'(R)-2f(R)+3\Box
f'(R)\Big](\rho_{m}+\rho_{rad})+\frac{5}{24M_{5}^{6}}(\rho_{m}+\rho_{rad})^{2}-R=0\,.
\end{equation}
In the next stage, we solve this equation for $\Box f'(R)$ to obtain
\begin{equation}
\Box
f'(R)=-\frac{1}{3}\Bigg[\Big(Rf'(R)-2f(R)\Big)+\frac{\rho_{m}+\rho_{rad}}{2M_{5}^{3}\,r_{c}}\Bigg]\pm
\frac{1}{r_{c}}\sqrt{\frac{2R}{15}}\,,
\end{equation}
Now we introduce an effective potential $V_{eff}$ which satisfies
$\Box f'(R)=\frac{\partial V_{eff}}{\partial f'(R)}\,$. This
effective potential has an extremum at
\begin{equation}
\Big[Rf'(R)-2f(R)\Big]+\frac{1}{m_{p}^{2}}(\rho_{m}+\rho_{rad})=\pm\frac{1}{r_{c}}\sqrt{\frac{6}{5}R}\,.
\end{equation}
In the high-curvature regime, where $f'(R)\simeq 1$ and
$\frac{f(R)}{R}\simeq 1$\,, we recover the standard DGP result (one
can compare this result with corresponding result obtained in Ref.
[18] to see the differences in this extended braneworld scenario)
\begin{equation}
R\pm\frac{1}{r_{c}}\sqrt{\frac{6}{5}R}=\frac{1}{m_{p}^{2}}\,(\rho_{m}+\rho_{rad})\,.
\end{equation}
The negative and positive signs in this equation are corresponding
to the DGP self-accelerating and normal branches respectively. In
which follows, we adopt the positive sign corresponding to the
\emph{normal branch} of the scenario. To investigate the expansion
history of the universe in this setup, we restrict ourselves to
those values of the model parameters that yield expansion histories
which are observationally viable. We note that the Hu-Sawicki $f(R)$
function introduced in Ref. [18], was interpreted as a cosmological
constant in the high-curvature regime. The motivation for that
interpretation was to obtain a $\Lambda$CDM behavior in the high
curvature (in comparison with $m^{2}$) regime. Here we show that in
a braneworld extension, the Hu-Sawicki induced gravity mimics the
$\Lambda$DGP model in the mentioned regime. As we have pointed out
previously, the phantom-like behavior can be realized from the
dynamical screening of the brane cosmological constant. In this
respect, we apply the same strategy to our model, so that the second
term in the Hu-Sawicki $f(R)$ function (that is, the second term in
the right hand side of equation (10)) mimics the role of an
effective cosmological constant on the DGP brane. Then this term
will be screened by $\frac{H}{\overline{r}_{c}}$ term in the late
time (see the normal branch of Eq. (3)).

In the case in which $R\gg m^{2}$\,, one can approximate Eq. (10) as
follows
\begin{equation}
\lim_{\frac{m^{2}}{R} \rightarrow 0}f(R)\approx
R-\frac{c_{1}}{c_{2}}m^{2}+\frac{c_{1}}{c_{2}^{2}}m^{2}\Big(\frac{R}{m^{2}}\Big)^{n}.
\end{equation}
During the late-time acceleration epoch, $f'_{0}(R)\simeq 1$ or
equivalently $R_{0}\gg m^{2}$ and we can apply the above
approximation. Also the curvature field is always near the minimum
of the effective potential. So, based on Eq. (19), we have
\begin{equation}
R+
\frac{1}{r_{c}}\sqrt{\frac{6}{5}R}=\frac{1}{m_{p}^{2}}\,(\rho_{m}+\rho_{rad})+2\frac{c_{1}}{c_{2}}m^{2}\,.
\end{equation}
Since $R$ in the $f(R)$ function is induced Ricci scalar on the
brane, we except crossover scale to affect on the constant
parameters $c_{1}$\,, $c_{2}$ and $m^{2}$. In Ref. [18] the authors
obtained $3m^{2}\equiv R_{c}=\frac{\rho_{0m}}{m_{p}^{2}}$ that
$\rho_{0m}$ is the present value of the matter density. But, in our
setup the present value of the matter density (see Eq. (20)) is
given by
\begin{equation}
R_{c}+0.9 \frac{\sqrt{R_{c}}}{r_{c}} =\frac{1}{m_{p}^{2}}\,(\rho_{
0m}+\rho_{ 0rad})\,.
\end{equation}
If we solve this equation for $R_{c}$, we find
\begin{equation}
3m^{2}\equiv
R_{c}\approx\Omega_{r_{c}}+3\Omega_{m}+3\Omega_{rad}\pm3\sqrt{\Omega_{r_{c}}\Big(0.068\Omega_{r_{c}}+\Omega_{m}+\Omega_{rad}\Big)}\,.
\end{equation}
Therefore, the DGP character of this extended modified gravity
scenario is addressed through $m^{2}$. As we have mentioned, at the
curvatures high compared with $m^{2}$, the second term on the right
hand side of equation (10) mimics the role of an effective
cosmological constant on the brane. In this respect, the second term
in the right hand side of equation (21) also mimics the role of a
cosmological constant on the brane in the high curvature regime.
With this motivation, we find
\begin{equation}
\frac{c_{1}}{c_{2}}\approx
\frac{18\Omega_{\Lambda}}{\Omega_{r_{c}}+3\Omega_{m}+3\Omega_{rad}\pm3\sqrt{\Omega_{r_{c}}\Big(0.068\Omega_{r_{c}}+\Omega_{m}+\Omega_{rad}\Big)}}\,.
\end{equation}
There is also a relation for $\frac{c_{1}}{c_{2}^{2}}$ as follows
\begin{equation}
\frac{c_{1}}{c_{2}^{2}}=\frac{1-f_{0}'(R)}{n}\bigg(\frac{R_{0}}{m^{2}}\bigg)^{n+1}\,,
\end{equation}
where $\frac{R_{0}}{m^{2}}$ in our setup can be calculated as
follows: firstly, by using Eqs. (22) and (25), we find
\begin{equation}
R+\frac{1}{r_{c}}\sqrt{\frac{6}{5}R}=\frac{1}{m_{p}^{2}}\,
(\rho_{_0m}a^{-3}+\rho_{0rad}\,a^{-4})+12\Omega_{\Lambda}\,,
\end{equation}
where $\rho_{_0m}$ can be omitted through Eq. (23) to obtain
\begin{equation}
R+\frac{1}{r_{c}}\sqrt{\frac{6}{5}R}=\bigg(3m^{2}+1.56\frac{m}{r_{c}}-\frac{1}{m_{p}^{2}}\,\rho_{0rad}\bigg)a^{-3}
+\frac{1}{m_{p}^{2}}\,\rho_{0rad}\,a^{-4}+12\Omega_{\Lambda}\,.
\end{equation}
Finally, if we solve this equation for $\sqrt{R}$\,, we find the
following relation for $\frac{R_{0}}{m^{2}}$
\begin{equation}
\frac{R_{0}}{m^{2}}\approx
\Bigg(-\frac{0.9\sqrt{\Omega_{r_{c}}}}{m}+\bigg[3\Big(1+\frac{0.52\sqrt{\Omega_{r_{c}}}}{m}\Big)^{2}+
\frac{12\Omega_{\Lambda}}{m^{2}}\bigg]^{1/2}\Bigg)^{2}\,,
\end{equation}
where $m$ is given by Eq. (24). Note that we have set $H_{0}$ and
$a(t_{0})$ equal to unity. These relations tell us that the free
parameters of this model are $n$, $\Omega_{m}$, $\Omega_{rad}$,
$\Omega_{r_{c}}$ and $f_{0}'(R)$, whereas the latter one is
constrained by the Solar-System tests. In fact, experimental data
show that $f'(R)-1~ <10^{-6}$\,, when $f'(R)$ is parameterized to be
exactly \,$1$ \,in the far past. To analyze the behavior of
$w_{curv}$\,, we specify the following ansatz for the scale factor
\begin{equation}
a(t)=\bigg(t^{2}+\frac{t_{0}}{1-\nu}\bigg)^{\frac{1}{1-\nu}}\,,
\end{equation}
where $\nu\neq 1$ is a free parameter [20]. By noting that the Ricci
scalar is $R=6(\frac{\ddot{a}}{a}+(\frac{\dot{a}}{a})^{2})$, one can
express the function $f(R)$ of equation (10) in terms of the
redshift $z$. Figure $2$ shows the variation of the effective
equation of state parameter versus the redshift. As we see in this
figure, in this class of models the \textit{curvature fluid} has an
effective phantom-like equation of state, $w_{curv} < -1$, at high
redshifts and then approaches the phantom divide ($w_{curv} = -1$)
at a redshift that increases by decreasing $n$.

\begin{figure}[htp]
\begin{center}\includegraphics{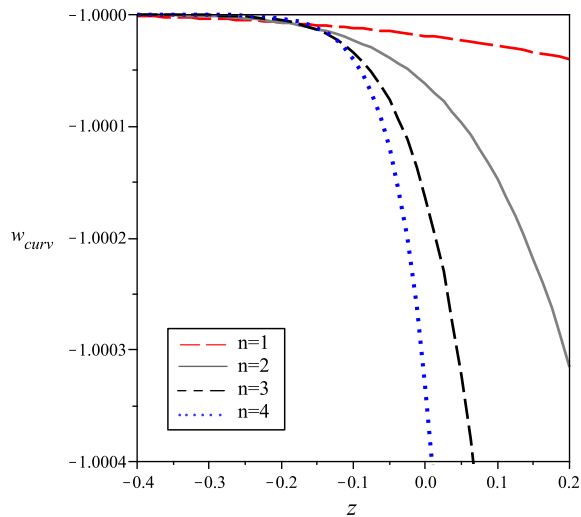} \vspace{4cm}
\end{center}
 \caption{\small {$w_{curv}$ versus the redshift for a Hu-Sawicki type
 modified induced gravity with $\Omega_{m}=0.27$,\, $\Omega_{rad}=0.3$,\,
 $\Omega_{\Lambda}=0.9$,\, $\Omega_{r_{c}}=0.01$
and $f_{0}'(R)-1\approx 10^{-6}$\,. As this figure shows, in this
class of models the \textit{curvature fluid} has an effective
phantom equation of state with $w_{curv} < -1$ at high redshifts.
This effective equation of state parameter approaches the phantom
divide line ($w_{curv}= -1$) at a redshift that increases by
decreasing the values of $n$. }}
\end{figure}

The main point here is that a modified induced gravity of the
Hu-Sawicki type in the DGP framework, gives an effective
phantom-like equation of state parameter for all values of $n$, and
all of these models approach asymptotically to the de Sitter phase
($w_{curv}=-1$). Therefore, the accelerated expansion of the
universe (the de Sitter phase) is necessarily a stable attractor
phase for this DGP-inspired $f(R)$ model. Based on the analysis
presented in the previous section within a phase space viewpoint and
also the outcomes of this section, we can conclude that a Hu-Sawicki
type modified induced gravity on the normal branch of the DGP setup
provides a cosmologically viable scenario. This is the case since it
contains a radiation dominated era followed by a matter dominated
era and finally a stable de Sitter phase in its expansion history.
In the next section we compare our model with observational data via
a cosmographic analysis. Our treatment here is mainly based on the
Ref. [29,30].

\section{Comparison with observational data}

While theoretical consistency of a physical theory is a primary
condition for viability of the theory, the observational consistency
of the model is necessary too. For this goal, in which follows we
discuss briefly observational status of our model via a cosmographic
analysis. Before that, we note that the DGP model is a testable
scenario with the same number of parameters as the standard
$\Lambda$CDM model, and has been constrained from many observational
data, such as the SNe Ia data set [21], the baryon mass fraction in
clusters of galaxies from the X-ray gas observation [22], CMB data
[23], the large scale structures [24] and the baryon acoustic
oscillation (BAO) peak [25], the observed Hubble parameter $H(z)$
data [26], the gravitational lensing surveys [27]. The observational
constraints on the DGP model with Gamma-ray bursts (GRBs) at high
redshift also obtained recently from the Union2 Type Ia supernovae
data set [28]. In [28] the authors are shown that by combining the
GRBs at high redshift with the Union2 data set, the WMAP7 results,
the BAO observation, the clusters' baryon mass fraction, and the
observed Hubble parameter data set and also in order to favor a flat
universe, the best fit of the density parameter values of the DGP
model are obtained as
$\{\Omega_{m},\Omega_{r_{c}}\}=\{0.235_{-0.014}^{+0.015}\,\,,0.138_{-0.048}^{+0.051}\}$
[28].

Here to compare our $f(R)$-DGP model with observational data we
adopt the cosmography approach. Cosmography approach is a useful
tool in order to constrain higher order gravity observationally
without need to solve field equations or addressing complicated
problems related to the growth of perturbations [29,30]. In this
case, one can define cosmographic parameters based on the fifth
order Taylor expansion of the scale factor. One can also relate the
characteristic quantities defining the $f(R)$-DGP model to the
mentioned cosmographic parameters. Therefore, a measurement of the
cosmographic parameters makes it possible to put constraints on
$f(R_{0})$ and its first three derivatives. The likelihood function
for the probe $s$ is defined as [31]
\begin{equation}
{\cal{L}}_{s}(\textbf{p})\propto\exp{(-\chi_{s}^{2}(\textbf{p})/2)}
\end{equation}
where
\begin{equation}
\chi_{s}^{2}(\textbf{p})=\sum_{n=1}^{{\cal{N}}_{s}}\frac{\Big[\mu_{obs}(z_{i})-\mu_{th}(z_{i},\textbf{p})\Big]^{2}}{\sigma_{\mu_{i}}(z_{i})}
\end{equation}
$\mu_{obs}(z_{i})$ are the observed distance modulus for the adopted
standard candle (such as SNe Ia) at the redshift $z_{i}$ with its
error $\sigma_{\mu_{i}}$.\, $\mu_{th}(z)$ are the theoretical values
of the distance modulus from cosmological models which read as
$\mu_{th}(z,\textbf{p})=25+5\log{D_{L}(z,\textbf{p})}$\, where
$D_{L}=H_{0}d_{L}$ is the luminosity distance. In the cosmography
approach, one can obtain an analytical expression for luminosity
distance versus the cosmographic parameters so that one require no
priori model to solve
$d_{L}=(1+z)\int_{_0}^{^z}\frac{dz'}{E(z')}$\,. By using the least
squares fitting that means the getting of $\chi_{s\, min}^{2}$, one
can obtain the suitable cosmographic parameters. In the next step,
one should relate the $f(R)$ function and its first three
derivatives to the cosmographic parameters to set constraints on the
parameters of the $f(R)$ function [29,30]. In this manner we
constrain observationally the parameters of a Hu-Sawicki type $f(R)$
induced gravity on the normal DGP brane by the cosmography approach.
Our strategy in this cosmographic approach is mainly based on the
recent paper by Bouhmadi-L\`{o}pez \emph{et al.} [30]. Firstly we
relate the functions $f(R_{0})$, $f'(R_{0})$, $f''(R_{0})$ and
$f'''(R_{0})$ to the parameters $R_{0}$, $\dot{R}_{0}$,
$\ddot{R}_{0}$,\, $(\frac{d^{3}R}{dt^{3}})_{0}$  and $\dot{H}_{0}$
which are expressed versus the cosmographic parameters by using the
Friedmann and Raychaudhuri equations at $t=t_{0}$\,. Now we have a
system of two equations with four unknowns. To expand the $f(R)$
function and its derivatives versus these cosmographic parameters,
we need to two further equations to close the system. In
4-dimensional $f(R)$ gravity, the Newtonian gravitational constant
$G$ is replaced by an effective (time dependent) quantity as
$G_{eff} = \frac{G}{f'(R)}$. On the other hand, it is reasonable to
assume that the present day value of $G_{eff}$ is the same as the
Newtonian one $G_{eff}(z=0)=G$ or $f'(R_{0})\simeq 1$\,. One may
note that the Hu-Sawicki model with this condition reduces to the
Einstein-Hilbert gravity with Lagrangian $f(R)=R$. In order to
resolve this problem, we can replace the condition $f'(R_{0})=1$
with $f'(R_{0})=(1+\epsilon)^{-1}$. Another relation can also be
obtained by differentiating the Raychaudhuri equation [29,30]. We
solve this system of four equations for four unknowns to obtain the
following relations
\begin{equation}
\left\{\begin{array}{ll}
\frac{f(R_{0})}{6H_{0}^{2}}=-\frac{{\cal{A}}_{0}\Omega_{m}+{\cal{B}}_{0}+{\cal{C}}_{0}(r_{c}H_{0})^{-1}}{\cal{D}}+\epsilon
(2-q_{0})\,,\vspace{0.5cm}\\
\frac{f''(R_{0})}{(6H_{0}^{2})^{-1}}=-\frac{{\cal{A}}_{2}\Omega_{m}+{\cal{B}}_{2}+{\cal{C}}_{2}(r_{c}H_{0})^{-1}}{\cal{D}}+\epsilon
\frac{{\cal{B}}_{2}-{\cal{C}}'_{2}}{\cal{D}}\,,\vspace{0.5cm}\\
\frac{f'''(R_{0})}{(6H_{0}^{2})^{-2}}=-\frac{{\cal{A}}_{3}\Omega_{m}+{\cal{B}}_{3}+{\cal{C}}_{3}(r_{c}H_{0})^{-1}}{(j_{0}-q_{0}-2)\cal{D}}+\epsilon
\frac{{\cal{B}}'_{3}-{\cal{C}}'_{3}}{(j_{0}-q_{0}-2)\cal{D}}\,,
\end{array}\right.
\end{equation}
where ${\cal{A}}_{i}$, ${\cal{B}}_{i}$, ${\cal{C}}_{i}$ and
$\cal{D}$ with $i=0,2,3$ are functions of $q_{0}$, $j_{0}$, $s_{0}$
and $l_{0}$ (these functions are defined in Ref. [30]). The new
quantities ${\cal{C}}'_{2}$, ${\cal{B}}'_{3}$, and ${\cal{C}}'_{3}$
are defined as follows
\begin{equation}
{\cal{C}}'_{2}=j_{0}(j_{0}-q_{0}^{2}-1)+q_{0}(q_{0}^{2}+q_{0}-3)-2\,,\quad
\end{equation}
$${\cal{B}}'_{3}=2j_{0}(2q_{0}^{2}+6q_{0}+j_{0}+3)+2q_{0}(15q_{0}^{2}+42q_{0}+39)$$
\begin{equation}
-2l_{0}(1+q_{0})-2s_{0}(q_{0}+j_{0})+24\,,\quad
\end{equation}
$${\cal{C}}'_{3}=j_{0}\big(-j_{0}+[2q_{0}+8]q_{0}+s_{0}+7\Big)+s_{0}(1-q_{0}^{2})$$
\begin{equation}
-q_{0}\Big(q_{0}[q_{0}^{2}+8q_{0}-2]-s_{0}-17\Big)+8\,.
\end{equation}
In the second step we have to determine the values of the
cosmographic parameters that have the best fit to the observational
data (by the least squares fitting). Instead, here we use a minimal
approach to parameterize the cosmographic parameters by the
phenomenological density parameters. In other words, the
cosmographic parameters will be calculated for a given
phenomenologically parameterized dark energy model. The best choice
is the $\Lambda$CDM model. In Ref. [30] the details of these
calculations are done. They finally obtained the following results
[30]
\begin{equation}
q_{0}=-1+\frac{3}{2}\Omega_{m}\,,\quad\quad j_{0}=1\,,\quad\quad
s_{0}=1-\frac{9}{2}\Omega_{m}\,,\quad\quad
l_{0}=1+3\Omega_{m}+\frac{27}{2}\Omega_{m}^{2}\,.
\end{equation}
Now one can substitute these results into equations (33) and
consider the observational conservative values $\Omega_{m}=0.266$
and $\Omega_{r_{c}}=10^{-4}$ where
$\Omega_{r_{c}}=(4r_{c}^{2}H^{2}_{0})^{-1}$ [2]. Finally, by
considering the first order approximation in $\epsilon$, one obtains
the following results
\begin{equation}
\frac{f(R_{0})}{6H_{0}^{2}}=0.849+2.6\epsilon\,,\quad
\frac{f''(R_{0})}{(6H_{0}^{2})^{-1}}=0.16-20.5\epsilon\,,\quad
\frac{f'''(R_{0})}{(6H_{0}^{2})^{-2}}=1.3+0.0176\epsilon
\end{equation}
In the HS model, there are four parameters $c_{1}$, $c_{2}$, $R_{c}$
and $n$ that can be constrained by observational data via the values
of the $f(R_{0})$ and its derivatives. So, we should create a system
of four equations in the four unknowns through equation (10) and its
first three derivatives. By solving equation (10) and its first
derivative for $c_{1}$ and $c_{2}$, with
$f(R_{0})=0.849+2.6\epsilon$ and $f'(R_{0})=(1+\epsilon)^{-1}$\,,
one finds [29]
\begin{equation}
\widetilde{c}_{1}\equiv
c_{1}R_{c}^{1-n}=\frac{n(1+\epsilon)}{\epsilon}R_{0}^{1-n}\bigg[1-\frac{0.849+2.6\epsilon}{R_{0}}\bigg]^{2}
\end{equation}
\begin{equation}
\widetilde{c}_{2}\equiv
c_{2}R_{c}^{-n}=\frac{n(1+\epsilon)}{\epsilon}R_{0}^{-n}\bigg[1-\frac{0.849+2.6\epsilon}{R_{0}}-\frac{\epsilon}{n(1+\epsilon)}\bigg]\,.
\end{equation}
By substituting relations (39) and (40) in HS $f(R)$ function and
its derivatives, it is obvious that parameter $R_{c}$ cancels out so
that we have to work with two parameters $\widetilde{c}_{1}$ and
$\widetilde{c}_{2}$ instead of three parameters $c_{1}$, $c_{2}$ and
$R_{c}$\,. In other words, $R_{c}$ cannot be obtained in this
fashion. By setting the second derivative of the HS function equal
to $f''(R_{0})=0.16-0.5\epsilon$, we get
\begin{equation}
n=\frac{[(0.849+2.6\epsilon)/R_{0}]+[(1+\epsilon)/\epsilon](1-[0.16-0.5\epsilon]/R_{0})-(1-\epsilon)/(1+\epsilon)}{1-(0.849+2.6\epsilon)/R_{0}}\,.
\end{equation}
In the last stage and in order to determine the value of
$\epsilon$\,, one can use the third derivative of the HS function
and setting $f'''(R_{0})=1.3+0.0176\epsilon$ to obtain the following
constraint (see also [29])
\begin{equation}
\frac{1.3+0.0176\epsilon}{0.849+2.6\epsilon}=\frac{1+\epsilon}{\epsilon}\frac{(0.16-0.5\epsilon)}{R_{0}}\bigg[R_{0}\frac{(0.16-0.5\epsilon)}{0.849+2.6\epsilon}
+\frac{\epsilon
(0.849+2.6\epsilon)}{1+\epsilon}\Big(1-\frac{2\epsilon}{1-(0.849+2.6\epsilon)/R_{0}}\Big)\bigg]
\end{equation}
Using this constrain, the acceptable value of $\epsilon$ is
$\simeq0.03$ (note that there are three other values of $\epsilon$
that are not acceptable since are very large). The value of $R_{0}$
is determined by $R_{0}=6H_{0}^{2}(1+q_{0})$ with
$q_{0}=-1+\frac{3}{2}\Omega_{m}$. By equation (41), we get
$n\simeq2$ and by equations (39) and (40) we obtain
$\widetilde{c}_{1}\simeq10$ and $\widetilde{c}_{2}\simeq0.7$\,. Note
that as we excepted these parameters are positive. The parameter
$R_{c}$ here plays the role of a scaling parameter. We obtain
$c_{1}$ and $c_{2}$ from equations (25) and (26) and then by using
their relation with $\widetilde{c}_{1}$ and $\widetilde{c}_{2}$, we
find $R_{c}\simeq0.018$\, which is a reasonable value for this
quantity.

\section{Summary}

Recently a mechanism to self-accelerate the normal branch of the DGP
model, which is known to be free from the ghost instabilities, has
been reported [17]. This mechanism is based on the modified induced
gravity. In this paper, firstly we studied the cosmological dynamics
of this model within a phase space approach. A de Sitter phase is
the simplest cosmological solution that exhibits acceleration. As we
have shown in a dynamical system viewpoint, this phase appears in
our generalized setup. In fact, based on the dynamical system
approach, we showed that there exists a de Sitter fixed point in
phase space of a general $f(R)$-DGP model. In order to investigate
the stability of this accelerating phase of expansion, we classified
the $f(R)$ functions in two different subspaces of the model
parameter space. We have shown that if the $f(R)$ induced gravity
plays effectively the role of a phantom scalar field in the
equivalent scalar-tensor theory, it leads to a stable de Sitter
solution and these models are cosmologically viable. Then, as an
specific model, we studied the Hu-Sawicki type modified induced
gravity in the DGP framework and we found that the equation of state
parameter of the curvature fluid has an effective phantom-like
character. The origin of the phantom-like behavior in the model
presented here can be due to the dynamical screening effect of the
curvature term (which plays effectively the role of a cosmological
constant in high-redshift regime on the brane). In other words, in
this case the phantom-like behavior has a pure gravitational origin.
We have shown also that the Hu-Sawicki modified induced gravity
mimics the $\Lambda$DGP model in the high-redshift regime. Since the
Hu-Sawicki modified induced gravity contains an early time radiation
dominated era followed by a matter domination era and then a stable
de Sitter phase in its expansion history, it is cosmologically a
viable scenario. This result is independent on the value of free
parameter $n$ of the Hu-Sawicki model. Finally we have tried to
constrain our model based on the observational data through a
cosmographic procedure. In this manner we obtained reasonable values
for parameters of the Hu-Sawicki induced gravity.\\

{\bf Acknowledgement}\\
We would like to thank an anonymous referee for insightful
suggestions.

\end{document}